\documentclass[9pt,epsf]{article}
\usepackage{CJKutf8}
\usepackage{amsmath}
\usepackage{booktabs, longtable}
\usepackage{caption2}
\usepackage{cite}
\usepackage{cmap}
\usepackage{enumerate}
\usepackage{fancybox}
\usepackage{flafter}
\usepackage{indentfirst}
\usepackage{latexsym}
\usepackage{listings}
\usepackage{pxfonts}
\usepackage{textcomp}
\usepackage{times}
\usepackage[T1]{fontenc}
\usepackage[dvips]{graphicx}
\usepackage{titletoc}
\usepackage{authblk}
\usepackage{multirow}
\usepackage{subfigure}
\usepackage{hyperref}
\usepackage{color}
\usepackage{graphicx}  
\usepackage{float}     
\usepackage{yhmath}
\usepackage{url}
\usepackage{appendix}
\usepackage{xcolor}
\usepackage{color}
\usepackage{flushend, cuted}
\usepackage{mathrsfs}

\newcommand{\apj}{APJ}

\newcommand{\mnras}{MNRAS}

\newcommand{\prd}{Phys. Rev. D}
\newcommand{\upcite}[1]{$^{\mbox{\scriptsize \cite{#1}}}$}

\definecolor{lightgray}{rgb}{.7,.7,.7}

\definecolor{red}{rgb}{1,0,0}

\definecolor{blue}{rgb}{0,0,1}

\begin{document}

\begin{center}
{\Large A mathematical form of force-free magnetosphere equation around Kerr black holes and its application to Meissner effect}
\vskip 1.0cm

\centerline {Xiao-Bo ~Gong $^{a,b,c,}$
\footnote{Corresponding authors at:Yunnan Observatory, Chinese Academy of Sciences, Kunming, 650011, China(X.-B Gong).  Department of Physics, South University of Science and Technology of China, Shenzhen, 518055, China(Y. Liao).
\\\emph{E-mail addresses:} gxbo@ynao.ac.cn(X.-B.Gong),~liaoy@mail.sustc.edu.cn(Y.Liao),~zyxu88@ynao.ac.cn(Z.-Y.Xu).},
~Yi Liao$^{d,~*}$,
~Zhao-Yi ~Xu$^{a,b,c}$}
 {\small $^a$Yunnan Observatory, Chinese Academy of Sciences, Kunming, 650011, China\\
 $^b$ Key Laboratory for the Structure and Evolution of Celestial Objects, Chinese Academy of Sciences, Kunming,
 650011, China\\
$^c$ University of Chinese Academy of Sciences, Beijing 100049, China\\
$^d$ Department of Physics, South University of Science and Technology of China, Shenzhen, 518055, China}
\end{center}
\vskip 1.5cm
\begin{abstract}
Based on the Lagrangian of the steady axisymmetric force-free magnetosphere (FFM) equation around Kerr black holes (KBHs),
we find that the FFM equation can be rewritten in a new form as
$f_{,rr}/(1-\mu^{2})+f_{,\mu\mu}/\Delta+K(f(r,\mu),r,\mu)=0$, where $\mu={\rm -\cos\theta}$.
With coordinate transformation, the above equation can be given as
$s_{,yy}+s_{,zz}+D(s(y,z),y,z)=0$. Using this form, we prove that the Meissner effect is not possessed by a KBH-FFM with the condition $d\omega/d A_{\phi}\leqslant0$ and $H_{\phi}(dH_{\phi}/dA_{\phi})\geqslant0$, here $A_{\phi}$ is the $\phi$ component of the vector potential $\vec{A}$, $\omega$ is the angular velocity of magnetic fields and ${H_{\phi}}$ corresponds to  twice the
poloidal electric current.
\end{abstract}

\newpage
{\section{Introduction}}
Blandford $\&$ Znajek (1977)\upcite{1} gave the equation for a force-free magnetosphere (FFM) of the curved Kerr spacetime, which can describe the energy extraction process. The rotational energy of the Kerr black hole (KBH) can be converted into the thermal and kinetic energy of the surrounding
 plasma, and there could exist an outgoing electromagnetically driven wind. In this physical picture the electron would emit many photons, which can produce a plentiful supply of the electron-positron pairs. The above energy source is a part of the central engines for gamma-ray bursts and active galactic nuclei. Exact analytic solutions of this highly
 nonlinear equation can provide a well understanding of the energy extraction process and can be helpful for numerical simulation.
 But there exists few analytic works to deal with this highly nonlinear equation. Based on a new mathematical form of this equation, we discuss about an interesting phenomenon called  Meissner effect, which is the expulsion of magnetic field lines out of
 the event horizon and the quenching of jet power for the KBH with high spin\upcite{7,1985MNRAS.212..899B}. 
 King et al. (1975)\upcite{10}  found this effect from the Wald vacuum solution\upcite{11}. Other types of black holes, such as Kerr-Newman black hole and string theory black holes, also show this behavior\upcite{2014PhRvD..90d3003P}.
 But this effect was never seen in the general relativistic magnetohydrodynamic simulations\upcite{6,7}. Penna(2014)\upcite{7} try to explain the absence of Meissner effect. They think it is a geometry effect and the steady axisymmetric fields only become
 radial near the event horizon to evade Meissner effect.
 but this explanation is not very satisfactory and convinced\upcite{6}. Pan $\&$ Yu (2016)\upcite{6} also try to answer this question in the steady axisymmetic force-free magnetosphere, but they need many
 special condition, such as, $H_{\phi}=\omega A_{\phi}$, where $A_{\phi}$ is the $\phi$ component of the vector potential $\vec{A}$, $\omega$ is the angular velocity of magnetic fields and ${H_{\phi}}$ corresponds to  twice the
poloidal electric current. In this paper, we try to answer this question using more general conditions.

 The paper is organized as follows.  In Section \ref{se:basic}, basic equation derived by Blandford $\&$ Znajek (1977)\upcite{1} and its Lagrangian are showed. Then we discuss about the physical meaning of its Lagrangian in Section \ref{se:phy}.
 In section \ref{se:equ}, we prove that this equation can be reduced to
 $\frac{f_{,rr}}{1-\mu^{2}}+\frac{f_{,\mu\mu}}{\Delta}+K(f,r,\mu)=0$ or $s_{,yy}+s_{,zz}+D(s,y,z)=0$. $K(f,r,\mu)$ and $D(s,y,z)$ are
 source functions. In Section \ref{se:Meissner}, we prove that the Meissner effect does not appear
  in a steady axisymmetric and magnetically dominated KBH-FFM with the condition
 $\frac{d\omega}{d A_{\phi}}\leqslant0$ and $\frac{dH_{\phi}^{2}}{dA_{\phi}}\geqslant0$. The summary is in Section \ref{se:con}.

{\section{Basic equation}\label{se:basic}}
The Kerr metric in Boyer-Lindquist coordinates is (with $c=G=1$)
\begin{eqnarray}
 ds^{2}&=&g_{\mu\nu}dx^{\mu}dx^{\nu} \nonumber
 \\ &=&-(1-\frac{2Mr}{\Sigma})dt^{2}-\frac{4Mar~{\rm \sin^{2}}\theta}{\Sigma}dtd\phi+\frac{\Sigma}{\Delta}dr^{2}
 +\Sigma d\theta^{2}+\frac{A~{\rm \sin^{2}}\theta}{\Sigma}d\phi^{2}.
\label{eq:metric}
\end{eqnarray}
where $x^{\nu}=(t,r,\theta,\phi), ~~ \nu=(0,1,2,3) \nonumber, \Delta=r^{2}-2Mr+a^{2}, \Sigma=r^{2}+a^{2}\cos^{2}\theta$ and
 $A=  (r^{2}+a^{2})\Sigma+2Mra^{2}\sin^{2}\theta=(r^{2}+a^{2})^{2}-\Delta a^{2}\sin^{2}\theta$.
Here, the notation $M$ is the KBH mass and $a$ its angular momentum.
The constraint differential equation for the FFM around KBHs is given by Menon $\&$ Dermer (2005)\upcite{4} as
\begin{equation}
\begin{aligned}
 \frac{\sqrt{\gamma}}{2\alpha\gamma_{\phi\phi}}\frac{dH_{\phi}^{2}}{dA_{\phi}}=&\omega\partial_{r}[\frac{\gamma_{\theta\theta}}{\alpha
 \sqrt{\gamma}}(\gamma_{\phi\phi}\omega+\beta_{\phi})A_{\phi,r}]+\omega\partial_{\theta}[\frac{\gamma_{rr}}{\alpha
 \sqrt{\gamma}}(\gamma_{\phi\phi}\omega+\beta_{\phi})A_{\phi,\theta}]     \\
 &+\partial_{r}[\frac{\gamma_{\theta\theta}}{\alpha\sqrt{\gamma}}(\beta^{2}-\alpha^{2}+\beta_{\phi}\omega)A_{\phi,r}]+
 \partial_{\theta}[\frac{\gamma_{rr}}{\alpha\sqrt{\gamma}}(\beta^{2}-\alpha^{2}+\beta_{\phi}\omega)A_{\phi,\theta}]
 \end{aligned}
\label{eq:constraint}
\end{equation}
where $\alpha=\sqrt{{\frac{\Delta\Sigma}{A}}}, \beta_{\phi}=-\frac{2Mra}{\Sigma}{\rm \sin^{2}}\theta,
\sqrt{\gamma}=\sqrt{\frac{A\Sigma}{\Delta}}{\rm \sin}\theta,
\gamma_{\phi\phi}=\frac{A{\rm \sin^{2}}\theta}{\Sigma}, \gamma_{\theta\theta}=\Sigma, \gamma_{rr}=\frac{\Sigma}{\Delta},
\beta^{2}-\alpha^{2}=\frac{2Mr}{\Sigma}-1$. $A_{\phi}$ is the $\phi$ component of the vector potential.
 The angular velocity $\omega(A_{\phi})$ is a function of $A_{\phi}$, this relation can be expressed as
 $d\omega=-\frac{1}{\Lambda}dA_{\phi}=\omega'dA_{\phi}$. ${H_{\phi}}(A_{\phi})$ corresponds to  twice the
poloidal electric current and it is also  a function of $A_{\phi}$.
 Use the Euler-Lagrange equation
 \begin{equation}
 \partial_{\nu}\frac{\partial\mathscr{L}}{\partial A_{\phi,\nu}}=\frac{\partial \mathscr{L}}{\partial A_{\phi}},
 \end{equation}
the Lagrangian for Eq.(\ref{eq:constraint}) can be expressed as
\begin{small}
\begin{equation}
\begin{aligned}
 \mathscr{L}&=\frac{\gamma_{\theta\theta}}{\alpha
 \sqrt{\gamma}}(\gamma_{\phi\phi}\omega^{2}+2\beta_{\phi}\omega+\beta^{2}-\alpha^{2})A_{\phi,r}^{2}+\frac{\gamma_{rr}}{\alpha
 \sqrt{\gamma}}(\gamma_{\phi\phi}\omega^{2}+2\beta_{\phi}\omega+\beta^{2}-\alpha^{2})A_{\phi,\theta}^{2}+
  \frac{\sqrt{\gamma}}{\alpha\gamma_{\phi\phi}}H_{\phi}^{2} \\
  &=\frac{1}{\sin \theta}(\frac{A~{\rm \sin^{2}}\theta}{\Sigma}\omega^{2}-\frac{4Mar~{\rm \sin^{2}}\theta}{\Sigma}\omega
  -1+\frac{2Mr}{\Sigma})(A_{\phi,r}^{2}+\frac{1}{\Delta}A_{\phi,\theta}^{2})+\frac{1}{\sin \theta}\frac{\Sigma}{\Delta}H_{\phi}^{2} \\
  &=\frac{1}{\sin \theta}(g_{33}\omega^{2}+2g_{03}\omega+g_{00})(A_{\phi,r}^{2}+\frac{1}{\Delta}A_{\phi,\theta}^{2})
  +\frac{1}{\sin \theta}g_{11}H_{\phi}^{2}.
   \end{aligned}
\label{eq:lagrangian}
\end{equation}
\end{small}
If $g_{33}\omega^{2}+2g_{03}\omega+g_{00}=0$, we can get the light surfaces, namely $(\omega-\Omega)\varpi=\pm\alpha$, where $\Omega=\frac{2aMr}{A}$,
  $\varpi=\sqrt{\frac{A}{\Sigma}}\sin\theta$. Let $\mu={\rm -\cos\theta}$, $L=g_{33}\omega^{2}+2g_{03}\omega+g_{00}$, and we assume that $L(A_{\phi},r,\mu)$ is a function of $A_{\phi},~r,~\mu$. Then Eq.(\ref{eq:constraint}) becomes
\begin{small}
\begin{equation}
\begin{aligned}
L(\frac{A_{\phi,rr}}{1-\mu^{2}}+\frac{A_{\phi,\mu\mu}}{\Delta})+
\frac{L{_{,r}}A_{\phi,r}}{1-\mu^{2}}+\frac{L{_{,\mu}}A_{\phi,\mu}}{\Delta}+
\frac{1}{2}L{_{,A_{\phi}}}(\frac{A_{\phi,r}^{2}}{1-\mu^{2}}+\frac{A_{\phi,\mu}^{2}}{\Delta})-
\frac{g_{11}}{1-\mu^{2}}H_{\phi}H_{\phi}'=0
\end{aligned}
\label{eq:gsequation}
\end{equation}
\end{small}
where $L{_{,r}}=g_{33,r}\omega^{2}+2g_{03,r}\omega+g_{00,r},~L{_{,\mu}}=g_{33,\mu}\omega^{2}+2g_{03,\mu}\omega+g_{00,\mu},
~L{_{,A_{\phi}}}=2g_{33}\omega\omega'+2g_{03}\omega', ~\omega'=\frac{d\omega}{d A_{\phi}},~ H_{\phi}'=\frac{d H_{\phi}}{d A_{\phi}}$.

{\section{Physical meaning of the Lagrangian}\label{se:phy}}
The Lorentz invariant ${\frac{1}{2}\rm \mathscr{F}^{2}}=\frac{1}{2}F_{\mu\nu}F^{\mu\nu}$ to the Carter observers is
$B^{2}-E^{2}$, where the Carter field components given by Znajek (1977)\upcite{8} are
\begin{equation}
\begin{aligned}
&E_{r}=[a-\omega(r^{2}+a^{2})]A_{\phi,r}/\Sigma,
\\&B_{r}=[(1-a\omega{\rm \sin^{2}}\theta)A_{\phi,\theta}]/(\Sigma{\rm \sin}\theta), \\
&E_{\theta}=[a-\omega(r^{2}+a^{2})]A_{\phi,\theta}/(\Sigma\sqrt{\Delta}),
\\&B_{\theta}=[-\sqrt{\Delta}(1-a\omega{\rm \sin^{2}}\theta)A_{\phi,r}]/(\Sigma{\rm \sin}\theta), \\
&E_{\phi}=0,
\\&B_{\phi}=H_{\phi}/(\sqrt{\Delta}{\rm \sin}\theta).
\end{aligned}
\label{eq:fieldcomponts}
\end{equation}
Then we can get
${\frac{1}{2}\rm \mathscr{F}^{2}}=(B_{r}^{2}+B_{\theta}^{2}+B_{\phi}^{2})-(E_{r}^{2}+E_{\theta}^{2}+E_{\phi}^{2})
=[1-(\nu_{c}^{\phi})^{2}][B_{r}^{2}+B_{\theta}^{2}]+B_{\phi}^{2}$, where $\nu_{c}^{\phi}$ is the
velocity of an observer rotating around the KBH with angular velocity $\omega$ with respect to
the Carter observers ($\nu_{c}^{\phi}=\frac{{\rm \sin}\theta}{\sqrt{\Delta}}\frac{\omega(r^{2}+a^{2})-a}{1-a\omega{\rm \sin^{2}}\theta}$).
The final result of the Lorentz invariant is
\begin{small}
\begin{equation}
\begin{aligned}
{\frac{1}{2} \rm \mathscr{F}^{2}}&=\frac{-1}{\Sigma\sin^{2} \theta}(\frac{A~{\rm \sin^{2}}\theta}{\Sigma}\omega^{2}-\frac{4Mar~{\rm \sin^{2}}\theta}{\Sigma}\omega
  -1+\frac{2Mr}{\Sigma})(A_{\phi,r}^{2}+\frac{1}{\Delta}A_{\phi,\theta}^{2})+\frac{1}{\sin^{2} \theta}\frac{1}{\Delta}H_{\phi}^{2} \\
  &=\frac{1}{\Sigma\sin \theta}(\frac{2}{\sin \theta}\frac{\Sigma}{\Delta}H_{\phi}^{2}-\mathscr{L})\\
  &=\frac{1}{\sqrt{-g}}(\frac{2}{\sin \theta}\frac{\Sigma}{\Delta}H_{\phi}^{2}-\mathscr{L}).
\end{aligned}
\label{eq:f2}
\end{equation}
\end{small}
Here, $g$ is the determinant of  the metric tensor ${g_{\mu\nu}}$. So $\mathscr{L}=-\sqrt{-g}[(B_{r}^{2}+B_{\theta}^{2}-B_{\phi}^{2})-(E_{r}^{2}+E_{\theta}^{2}+E_{\phi}^{2})]$. This relation can
be found in MacDonald $\&$ Thorne (1982)\upcite{3}.

 The inner and the outer light surface will never intersect each other\upcite{2} when $0< \omega <\frac{a}{2Mr_{+}}$,
there exists a region between the inner light surface and the outer light surface such that $L < 0$. According to the equation: $[\omega(r^{2}+a^{2})-a]^{2}{\rm \sin^{2}}-\Delta(1-a\omega{\rm \sin^{2}}\theta)^{2}=\Sigma L$, there are three cases: (i). $L<0$ implies $|\nu_{c}^{\phi}|<1$, which means $L$ is time-like. (ii). $L>0 $ leads to $|\nu_{c}^{\phi}|>1$, it will be space-like. (iii). $L=0$ leads to $|\nu_{c}^{\phi}|=1$, it will be null\upcite{3}. Inside the inner light surface we have $[1-(\nu_{c}^{\phi})^{2}][B_{r}^{2}+B_{\theta}^{2}] <0 $, then ${\frac{1}{2}\rm \mathscr{F}^{2}} >0 $ reduces to $B_{\phi}^{2}>0 $ and $H_{\phi}^{2} >0$. Komissarov (2004)\upcite{2} analyzed this poloidal electric field in detail at this region.

{\section{Equation forms}\label{se:equ}}
Let $p\equiv A_{\phi}$ and $f(r,\mu)=F(p,r,\mu)$. Here, $F(p,r,\mu)$ and $L(p,r,\mu)$ are two
 functions of $p, r$ and $\mu$. $f(r,\mu)$ is a function of  the variables $r$ and $\mu$. Then we have the following results:
$f_{,rr}=p_{,rr}F_{,p}+p_{,r}^{2}F_{,pp}+2p_{,r}F_{,pr}+F_{,rr}, ~
f_{,\mu\mu}=p_{,\mu\mu}F_{,p}+p_{,\mu}^{2}F_{,pp}+2p_{,r}F_{,p\mu}+F_{,\mu\mu}$.
We apply the above results to compare  Eq.(\ref{eq:gsequation}) with equation
 $\frac{f_{,rr}}{1-\mu^{2}}+\frac{f_{,\mu\mu}}{\Delta}+K(f,r,\mu)=0$
and find that
\begin{equation}
\begin{aligned}
\frac{F_{,p}}{L}=\frac{2F_{,pp}}{L_{,p}}=
\frac{2F_{,pr}}{L_{,r}}=\frac{2F_{,p\mu}}{L_{,\mu}},
\end{aligned}
\label{eq:fg22}
\end{equation}
 then we have $F_{,p}=\sqrt{\left|L\right|}$. If we know the relation between $\omega$ and $A_{\phi}$, we can also get the relation between $f$ and $A_{\phi}$ through $F=\int \sqrt{\left|L\right|}dA_{\phi}$. If ${\omega=\rm constant}$, then $F(A_{\phi},r,\mu)=\sqrt{\left|L\right|}A_{\phi}$.
Apply Eq.(\ref{eq:fg22}) to Eq.(\ref{eq:gsequation}) gives
\begin{equation}
\frac{f_{,rr}}{1-\mu^{2}}+\frac{f_{,\mu\mu}}{\Delta}
-(\frac{F_{,rr}}{1-\mu^{2}}+\frac{F_{,\mu\mu}}{\Delta})-
\frac{F_{,p}g_{11}}{L(1-\mu^{2})}H_{\phi}H_{\phi}'=0.
\label{eq:fFF22}
\end{equation}
 $F_{,rr}$ and $F_{,\mu\mu}$ are not partial derivatives with respect to $p$, so
 $F_{,rr}$ and $F_{,\mu\mu}$ have no terms of $p_{,r}, p_{,rr}, p_{,\mu}, p_{,\mu\mu}$, then $C=-(\frac{F_{,rr}}{1-\mu^{2}}+\frac{F_{,\mu\mu}}{\Delta})-\frac{F_{,p}g_{11}}{L(1-\mu^{2})}H_{\phi}H_{\phi}'$. $C$  is just a function of $p, r, \mu$. On the other hand, $f(r,\mu)=F(p,r,\mu)$,
 we have $p=p(f,r,\mu)$ so that $C(p,r,\mu)=C(p(f,r,\mu),r,\mu)=K(f,r,\mu)$.

\begin{figure}
 \centering
 \includegraphics[angle=0,scale=0.526,bbllx=151pt,bblly=223pt,bburx=460pt,bbury=569pt]{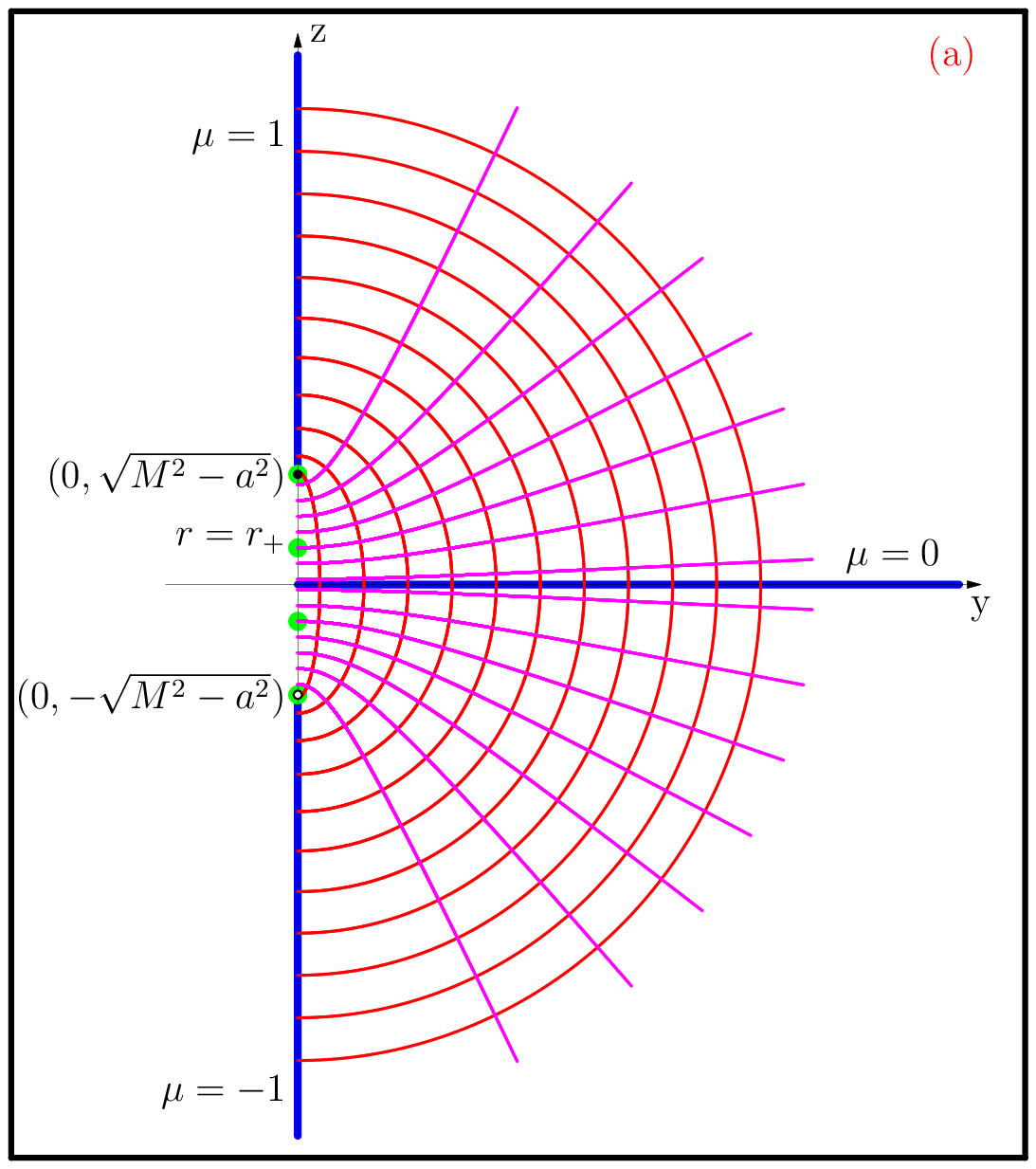}
  \includegraphics[angle=0,scale=0.526,bbllx=132pt,bblly=222pt,bburx=474pt,bbury=568pt]{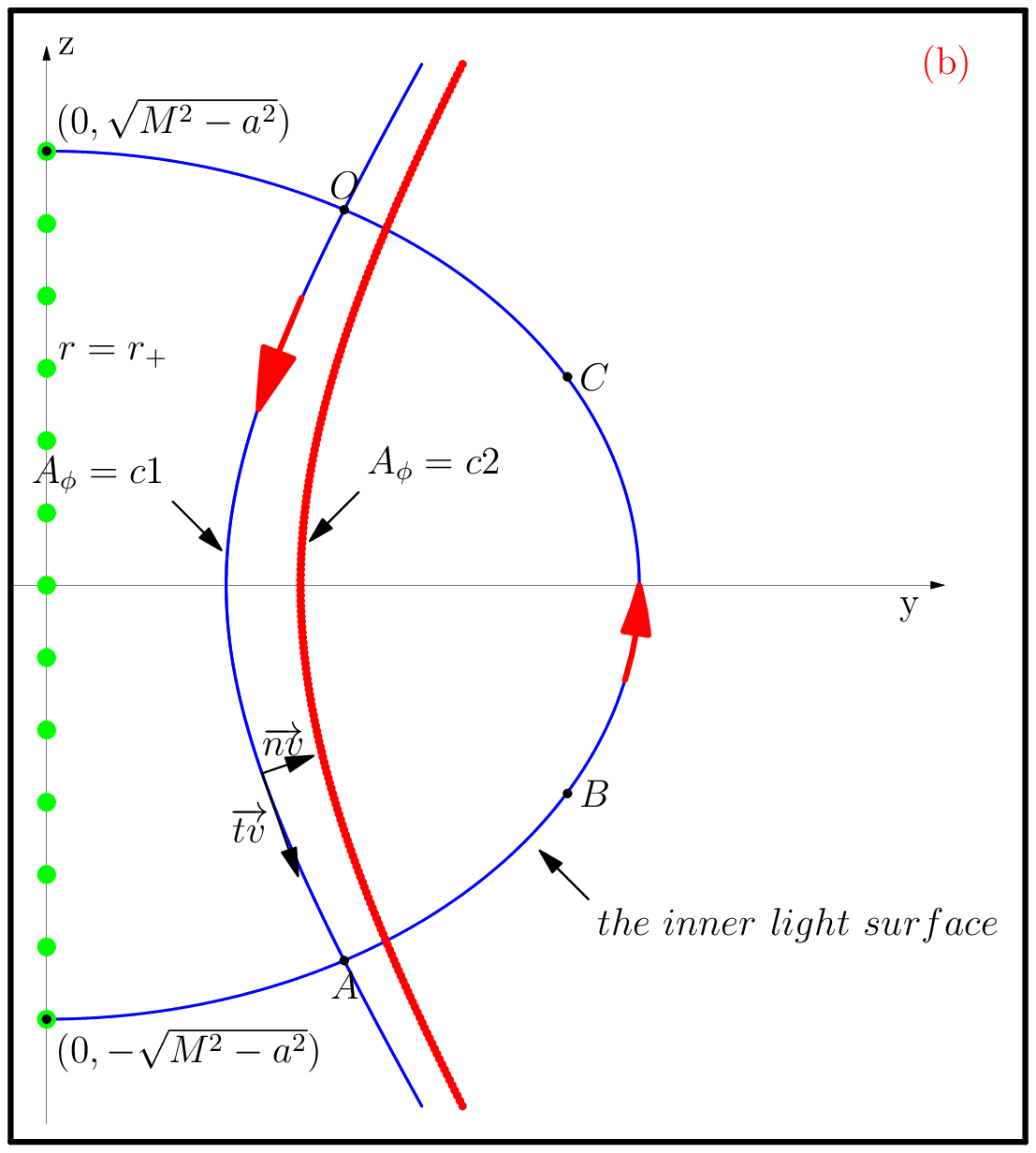}
    \caption{(a): coordinate transformation. Three blue thick solid lines represent $\mu=1,~ \mu=0,~ \mu=-1$, respectively.
    The  ellipses are $\frac{z^{2}}{(r-M)^{2}}+\frac{y^{2}}{\Delta}=1$ with $r= constant$, the hyperbolas are $\frac{z^{2}}{\mu^{2}(M^{2}-a^{2})}-\frac{y^{2}}{(1-\mu^{2})(M^{2}-a^{2})}=1$ with $\mu= constant$. (b): the inner light surface $\wideparen{ABCO}$ and the configuration of magnetic fields. $c1, c2$ are constant.
     }
  \label{Fig:Mei}
\end{figure}

Using coordinate transformation, we can simplify $\frac{\zeta_{,rr}}{1-\mu^{2}}+\frac{\zeta_{,\mu\mu}}{\Delta}$. At first,
\begin{equation}
\begin{aligned}
\frac{\zeta_{,rr}}{1-\mu^{2}}+\frac{\zeta_{,\mu\mu}}{\Delta}=
\begin{cases}\frac{1}{M^{2}-a^{2}}( \frac{\zeta_{,RR}}{1-\mu^{2}}+\frac{\zeta_{,\mu\mu}}{R^{2}-1}),~
&{\rm with}~ R=\frac{r-M}{\sqrt{M^{2}-a^{2}}},~{\rm if} ~ a^{2}<M^{2} \\
\frac{\zeta_{,RR}}{1-\mu^{2}}+\frac{\zeta_{,\mu\mu}}{R^{2}}, ~
&{\rm with}  ~R=r-M,~~~~ {\rm if} ~ ~ a^{2}=M^{2}
\end{cases}
\end{aligned}
\label{eq:gelin24}
\end{equation}
Now we transform the equation$\frac{f_{,RR}}{1-\mu^{2}}+\frac{f_{,\mu\mu}}{R^{2}-1}=0$ in $(R, \mu)$ space to $(\overline{y},\overline{z})$ space, and
\begin{equation}
\begin{aligned}
&\overline{y}=\overline{y}(R, \mu), ~~\overline{z}=\overline{z}(R, \mu) \\
\frac{f_{,RR}}{1-\mu^{2}}+\frac{f_{,\mu\mu}}{R^{2}-1}&=a_{11}f_{,\overline{y}~\overline{y}}+2a_{12}f_{,\overline{y}~\overline{z}}
+a_{22}f_{,\overline{z}~\overline{z}}+
b_{1}f_{,\overline{y}}+b_{2}f_{,\overline{z}} \\
\end{aligned}
\label{eq:transform}
\end{equation}
where, $a_{11}=\frac{\overline{y}_{,R}^{2}}{1-\mu^{2}}+\frac{\overline{y}_{,\mu}^{2}}{R^{2}-1},~
a_{22}=\frac{\overline{z}_{,R}^{2}}{1-\mu^{2}}+\frac{\overline{z}_{,\mu}^{2}}{R^{2}-1},~
a_{12}=\frac{\overline{y}_{,R}\overline{z}_{,R}}{1-\mu^{2}}+\frac{\overline{y}_{,\mu}\overline{z}_{,\mu}}{R^{2}-1},~
b_{1}=\frac{\overline{y}_{,RR}}{1-\mu^{2}}+\frac{\overline{y}_{,\mu\mu}}{R^{2}-1},~
b_{2}=\frac{\overline{z}_{,RR}}{1-\mu^{2}}+\frac{\overline{z}_{,\mu\mu}}{R^{2}-1}.$
Solving the equations $a_{11}=a_{22}~{\rm and}~ a_{12}=0$, we can obtain the following equations:
$\overline{y}_{,R}=\sqrt{\frac{1-\mu^{2}}{R^{2}-1}}\overline{z}_{,\mu}, ~
 \overline{y}_{,\mu}=-\sqrt{\frac{R^{2}-1}{1-\mu^{2}}}\overline{z}_{,R}$.
The relation $\overline{y}_{,R\mu}=\overline{y}_{,\mu R}$ leads to
\begin{equation}
\frac{\overline{z}_{,RR}}{1-\mu^{2}}+\frac{\overline{z}_{,\mu\mu}}{R^{2}-1}+\frac{R~ \overline{z}_{,R}}{(1-\mu^{2})(R^{2}-1)}
-\frac{\mu \overline{z}_{,\mu}}{(1-\mu^{2})(R^{2}-1)}=0
\label{eq:transformeq}
\end{equation}
The substitution $\overline{z}=\overline{R}(R)U(\mu)$ will separate Eq.(\ref{eq:transformeq}) into
\begin{equation}
\begin{aligned}
(1-\mu^{2})U_{,\mu\mu}-\mu U_{,\mu}+n^{2}U=0, ~~~
(1-R^{2})\overline{R}_{,RR}-R \overline{R}_{,R}+n^{2}\overline{R}=0,
\end{aligned}
\label{eq:transformeq2}
\end{equation}
where $n$ is a non-negative integer.  The solutions of Eq.(\ref{eq:transformeq2}) are the Chebyshev polynomials of the first kind.
The simplest solution of Eq.(\ref{eq:transformeq2}) is $\overline{z}=\mu R ~{\rm so~ that}~ \overline{y}=\sqrt{(1-\mu^{2})(R^{2}-1)} $. $\frac{f_{,RR}}{1-\mu^{2}}+\frac{f_{,\mu\mu}}{R^{2}-1}=0$
takes the form $(\frac{\mu^{2}}{1-\mu^{2}}+\frac{R^{2}}{R^{2}-1})(f_{,\overline{y}~\overline{y}}+f_{,\overline{z}~\overline{z}}-
\frac{f_{,\overline{y}}}{\overline{y}})=0$. Similarly,
$\frac{f_{,RR}}{1-\mu^{2}}+\frac{f_{,\mu\mu}}{R^{2}}=0$ can be transformed into the form $\frac{1}{1-\mu^{2}}(f_{,\overline{y}~\overline{y}}+f_{,\overline{z}~\overline{z}}-\frac{f_{,\overline{y}}}{\overline{y}})=0$ with coordinate transformation $\overline{z}=\mu R~{\rm and}~ \overline{y}=\sqrt{1-\mu^{2}}R$.
Let $f=\overline{s}\sqrt{\overline{y}}$, then $f_{,\overline{y}~\overline{y}}+f_{,\overline{z}~\overline{z}}-\frac{f_{,\overline{y}}}{\overline{y}}=
\sqrt{\overline{y}}(\overline{s}_{,\overline{y}~\overline{y}}
+\overline{s}_{,\overline{z}~\overline{z}}-\frac{3\overline{s}}{4\overline{y}^{^{2}}})$.
Finally, Eq.(\ref{eq:fFF22}) becomes
\begin{equation}
\begin{aligned}
s_{,yy}+s_{,zz}=\frac{3s}{4y^{^{2}}}+
\frac{(1-\mu^{2})\Delta}{\sqrt{y}[\Delta+(m^{2}-a^{2})(1-\mu^{2})]}[(\frac{F_{,rr}}{1-\mu^{2}}+\frac{F_{,\mu\mu}}{\Delta})+
\frac{F_{,p}g_{11}}{L(1-\mu^{2})}H_{\phi}H_{\phi}']  \\
\end{aligned}
\label{eq:transformeqfinall}
\end{equation}
with variable transformation $z=\mu(r-M),~ y=\sqrt{(1-\mu^{2})\Delta} ~ {\rm and}~ f=s\sqrt{y}$ (see Fig.\ref{Fig:Mei} (a)).
So $D(s,y,z)=-\frac{3s}{4y^{^{2}}}-
\frac{(1-\mu^{2})\Delta}{\sqrt{y}[\Delta+(m^{2}-a^{2})(1-\mu^{2})]}[(\frac{F_{,rr}}{1-\mu^{2}}+\frac{F_{,\mu\mu}}{\Delta})+
\frac{F_{,p}g_{11}}{L(1-\mu^{2})}H_{\phi}H_{\phi}']$.

We can make use of the following relations to re-express $\frac{F_{,rr}}{1-\mu^{2}}+\frac{F_{,\mu\mu}}{\Delta}$.
Because $F_{,rr}=\int F_{,p}[\frac{L_{,rr}}{2L}-(\frac{L_{,r}}{2L})^{2}]dx,~
F_{,\mu\mu}=\int F_{,p}[\frac{L_{,\mu\mu}}{2L}-(\frac{L_{,\mu}}{2L})^{2}]dx,~
L_{,rr}=2(1-\mu^{2})\omega^{2}+[a(1-\mu^{2})\omega-1]^{2}g_{00,rr},~
L_{,\mu\mu}=-2\Delta\omega^{2}+[\frac{r^{2}+a^{2}}{a}\omega-1]^{2}g_{00,\mu\mu},~ g_{00,\mu\mu}=-a^{2}g_{00,rr},$
we can obtain
\begin{equation}
\frac{L_{,rr}}{1-\mu^{2}}+\frac{L_{,\mu\mu}}{\Delta}=\{\frac{[a(1-\mu^{2})\omega-1]^{2}}{1-\mu^{2}}-\frac{[ (r^{2}+a^{2})\omega-a]^{2}}{\Delta}\}g_{00,rr}=-\frac{g_{00,rr}\Sigma  L}{\Delta(1-\mu^{2})},
\end{equation}
and Eq.(\ref{eq:fFF22}) becomes
\begin{equation}
\frac{f_{,rr}}{1-\mu^{2}}+\frac{f_{,\mu\mu}}{\Delta}+\int\frac{F_{,p}}{4L^{2}}(\frac{L_{,r}^{2}}{1-\mu^{2}}+\frac{L_{,\mu}^{2}}{\Delta})dp
+\frac{g_{00,rr}f}{2\Delta(1-\mu^{2})}-\frac{F_{,p}\Sigma}{L\Delta(1-\mu^{2})}H_{\phi}H_{\phi}'=0.
\label{eq:f22ll}
\end{equation}
$L_{,\mu}^{2}$ is positive for $r > r_{+}=M+\sqrt{M^{2}-a^{2}}$ and $\mu\neq 0$ so that $\frac{F_{,p}}{4L^{2}}(\frac{L_{,r}^{2}}{1-\mu^{2}}+\frac{L_{,\mu}^{2}}{\Delta})$ is singular at the light surfaces.

{\section{Meissner effect}\label{se:Meissner}}
 Our proof is similar to that in Pan $\&$ Yu (2016)\upcite{6}. But  theirs needs many hypotheses and only applies to a special case. If there is Meissner effect, then all magnetic field lines should not cross the event horizon. Because
 $\overline{B}_{r}=F_{\theta\phi}=A_{\phi,\theta}, ~\overline{B}_{\theta}=F_{\phi r}=-A_{\phi,r}$. the tangential vector of $A_{\phi}=constant$ is $(\overline{B}_{r},~\overline{B}_{\theta})$ in $(r,~ \theta)$ space. If there exists a magnetic field line which only cross the inner light surface and does not cross the event horizon, then there must exists a curve $A_{\phi}=constant$
 which only cross the inner light surface and does not cross the event horizon.
 We will prove that magnetic field lines of this type do not exist, so
 that  the curves like $\wideparen{OA}$ in the Fig.\ref{Fig:Mei} (b) ,  do not exist. And we do not require the FFM to be symmetric with respect to the equator.
In $(y,z)$  space, Eq.(\ref{eq:gsequation}) becomes
\begin{equation}
\begin{aligned}
L(A_{\phi,yy}+&A_{\phi,zz}-\frac{1}{y}A_{\phi,y})+L_{,y}A_{\phi,y}+L_{,z}A_{\phi,z}+\frac{1}{2}L_{,A_{\phi}}(A_{\phi,y}^{2}+A_{\phi,z}^{2})=\kappa H_{\phi}H_{\phi}'~~~{\rm or} ~~~\\
& y\nabla_{2}\cdot(\frac{L}{y}\nabla_{2} A_{\phi})=\frac{1}{2}L_{,A_{\phi}}(A_{\phi,y}^{2}+A_{\phi,z}^{2})+\kappa H_{\phi}H_{\phi}',
\end{aligned}
\label{eq:tuo}
\end{equation}
 where $\kappa$ is $\frac{\Sigma}{\Delta+(m^{2}-a^{2})(1-\mu^{2})}$.
 Making use of Green's theorem and Eq.(\ref{eq:tuo}), we obtain
 \begin{footnotesize}
\begin{equation}
\begin{aligned}
\int\int\nabla_{2}\cdot(\frac{L}{y}\nabla_{2} A_{\phi})dydz=\int\int\frac{1}{y}[\frac{1}{2}L_{,A_{\phi}}(A_{\phi,y}^{2}+A_{\phi,z}^{2})+\kappa H_{\phi}H_{\phi}']dydz=\oint\limits_{\wideparen{OABCO}}\frac{L}{y}(A_{\phi,y}dz-A_{\phi,z}dy).
\end{aligned}
\label{eq:gel2}
\end{equation}
\end{footnotesize}
In Fig.\ref{Fig:Mei} (b)  the tangential vector and the normal vector of $\wideparen{OA}$ is $\overrightarrow{tv}=(dy,~ dz)$ and
  $\overrightarrow{nv}=(A_{\phi,y}, A_{\phi,z})$, respectively. They have the relation $A_{\phi,y}dy+A_{\phi,z}dz=0$. The curve $\wideparen{ABCO}$ is the inner light surface, and $\int\limits_{\wideparen{ABCO}}\frac{L}{y}(A_{\phi,y}dz-A_{\phi,z}dy)=0$. Using boundary condition $A_{\phi,y}(r=\infty) >0$, we can make sure that there must exists a region where the of $A_{\phi}$ on the right equipotential line is higher than  that on the left (.i.e. $c2>c1$). This boundary condition is a necessary condition for the extraction of energy. Then in Fig.\ref{Fig:Mei} (b) the cross product $\overrightarrow{tv}\times\overrightarrow{nv}$ is always perpendicular to the paper plane and pointing outside so that $-A_{\phi,y}dz+A_{\phi,z}dy >0 $. Inside the inner light surface $\frac{L}{y}>0$ leads to
$\int\limits_{\wideparen{OA}}\frac{L}{y}(A_{\phi,y}dz-A_{\phi,z}dy)<0$, and $g_{33}\omega+g_{03}$ is less than $0$ under the condition $0< \omega <\frac{a}{2Mr_{+}}$\upcite{2}. $\kappa$ is always bigger than $0$. If we assume that $H_{\phi}H_{\phi}'\geqslant0$ and $\omega'\leqslant0$, then $L_{,A_{\phi}}=2(g_{33}\omega+g_{03})\omega' \geqslant0$, and  $\int\int\frac{1}{y}(\frac{1}{2}L_{,A_{\phi}}(A_{\phi,y}^{2}+A_{\phi,z}^{2})+\kappa H_{\phi}H_{\phi}')dydz\geqslant0$, which is in
contradiction to $\oint\limits_{\wideparen{OABCO}}\frac{L}{y}(A_{\phi,y}dz-A_{\phi,z}dy) < 0$(see Eq.(\ref{eq:gel2})).
The inner light surface is located between event horizon and the the ergosphere\upcite{2}. So the magnetic field lines which cross the inner light surface must also cross the event horizon. The Meissner effect expels magnetic fields out of the event horizon and
it also  expels magnetic fields out of the inner light surface. The magnetic fields outside the inner light surface could not
  go to the region between the event horizon and the inner light surface, and the Meissner effect expels magnetic fields out of the event horizon, so there can only exist closed poloidal field line or
constant  $A_{\phi}$, $\omega$ and $H_{\phi}$ between the event horizon and the inner light surface. But the
closed poloidal fields does not exist. The reason is as follows.
 One of the necessary conditions for the steady-state force-free approximation not
 break down is ${\frac{1}{2}\rm \mathscr{F}^{2}}>0$\upcite{8,2}[see Eq. (71), Eq.(72) and Section 3 in Komissarov (2004)\upcite{2}].
Because Gralla $\&$ Jacobson (2014)\upcite{9} have proved:

\emph{a stationary, axisymmetric, force-free and magnetically dominated field (${\frac{1}{2}\rm \mathscr{F}^{2}}>0$)  configuration cannot possess  a closed loop of poloidal field line} [see Eq. (98) in Gralla  $\&$ Jacobson (2014)\upcite{9}].\\
This leads to any closed poloidal field line not exist in the region between event horizon and the inner light surface. The final case left is  that $A_{\phi}$, $\omega$ and $H_{\phi}$ must be  constant in  this region .  The boundary condition $H_{\phi}=\frac{{\rm \sin}\theta(\omega(r_{+}^{2}+a^{2})-a)}{r_{+}^{2}+a^{2}\cos^{2}\theta}A_{\phi,\theta}$\upcite{2} at the event horizon leads to  $H_{\phi}=0$ in  the region between the event horizon and the inner light surface, so no magnetic field or electric field exists.  This means ${\frac{1}{2}\rm \mathscr{F}^{2}}=0$ in  this region. These configuration are unrealistic.
   So the Meissner effect does not appear in a stationary, axisymmetric and magnetically dominated  KBH-FFM.
If $\omega= constant$, the relations between $H_{\phi}$ and $A_{\phi}$ in Pan $\&$ Yu (2016)\upcite{6},  which include the Blandfold-Znajek
monopole solution,  are consistent with the condition $\frac{d\omega}{d A_{\phi}}\leqslant0$ and $\frac{dH_{\phi}^{2}}{dA_{\phi}}\geqslant0$.
The numerical simulation in Nathanail $\&$ Contopoulos (2014)\upcite{5} also shows that no Meissner effect occurs for some magnetospheres, which are consistent with the condition $\frac{d\omega}{d A_{\phi}}\leqslant0$ and $\frac{dH_{\phi}^{2}}{dA_{\phi}}\geqslant0$. When $\omega=\frac{a}{2Mr_{+}}$, the inner light surface will meet the event horizon, there will be no rotational energy for the KBH to extract\upcite{1}, and our proof do not apply to this situation. We think the Meissner effect will occur in
 this case.

{\section{Conclusion}\label{se:con}}
 We give a new mathematical form of FFM equation. The simplest case of this equation have the form $s_{,yy}+s_{,zz}=\rho s$, where $\rho$ and $s$ are independent. The stable Schr\"{o}dinger equation has this form
 and it has some analytic solutions. However in our case, $\rho$  is a
 very complicated function of $y,z$ so it is hard to give an analytic solution. Although the Green's function of Laplace's equation can be easily obtained for the positive quarter-plane or half-plane, the integral equation is hard to solve.  In other cases where FFM equations are high nonlinear, the solutions are more difficult to get. On the other hand, analyzing the light surface function
 $L=g_{33}\omega^{2}+2g_{03}\omega+g_{00}$ provides important insights on understanding this equation. We will focus on it in our future work.

{\section*{Acknowledgments}}
This work was partly supported by National Natural Science Foundation of China (NSFC) under grants (11373063, 11273053, 11390374, 11573061
 and XDB09010202)
and Yunnan Foundation (grant No. 2011CI053).

\end{document}